\documentclass[runningheads]{llncs}
\usepackage[T1]{fontenc}
\usepackage{graphicx}
\usepackage{colortbl}
\usepackage{multirow}
\definecolor{mygray}{gray}{.85}
\usepackage{booktabs}
\usepackage{amsmath} 
\usepackage{amssymb}
\begin{document}
\title{Self-supervised Hierarchical Representation for Medication Recommendation
}
%
%
\author{Yuliang Liang\inst{1}
\and
Yuting Liu\inst{1} \and
Yizhou Dang\inst{1} \and
Enneng Yang\inst{1} \and
Guibing Guo\inst{1} \and
Wei Cai\inst{2} \and
Jianzhe Zhao\inst{1} \and
Xingwei Wang\inst{1} 
}

%
\institute{Northeastern University, China \and Neusoft Research of Intelligent Healthcare Technology Co. Ltd., China
}
\maketitle              
\begin{abstract}
Medication recommender is to suggest appropriate medication combinations based on a patient's health history, e.g., diagnoses and procedures. Existing works represent different diagnoses/procedures well separated by one-hot encodings. However, they ignore the latent hierarchical structures of these medical terms, undermining the generalization performance of the model.
For example, ``Respiratory Diseases'', ``Chronic Respiratory Diseases'' and ``Chronic Bronchiti'' have a hierarchical relationship, progressing from general to specific. To address this issue, we propose a novel \underline{hier}archical encoder named HIER to hierarchically represent diagnoses and procedures, which is based on standard medical codes and compatible with any existing methods.
Specifically, the proposed method learns relation embedding with a self-supervised objective for incorporating the neighbor hierarchical structure. Additionally, we develop the position encoding to explicitly introduce global hierarchical position. Extensive experiments demonstrate significant and consistent improvements in recommendation accuracy across four baselines and two real-world clinical datasets. 


\end{abstract}
\vspace{-30pt}
\section{Introduction}
Medication recommendation systems understand patients' health status from electronic health records (EHRs) and tailor a safe medication combination. In EHRs, each patient's medical history is documented as a series of visits, typically involving diagnoses, procedures and medications. Existing works often represent diagnoses and procedures with one-hot encodings. The model then learns their embeddings in a low-dimensional space from patient data~\cite{safedrug,wang2022hienet,yang2023molerec}.
This embedding method organizes symbolic objects (e.g., words, entities and concepts) and has proven successful in numerous applications~\cite{devlin2018bert,dosovitskiy2020vit,he2020lightgcn}.
In healthcare, a universal language—namely, a standard coding system—is used to categorize diseases and conditions hierarchically.
This hierarchical structure enables constraints to be imposed on the embedding space, allowing the model to effectively learn commonalities within each disease category.


Recent studies in medication recommendation mainly focus on integrating drug-related knowledge into the models, such as drug-drug interaction graphs~\cite{www2021dpr,tois2023dpr}, molecular structures of drugs~\cite{safedrug,yang2023molerec}, and medical knowledge graphs linking diseases with medications~\cite{tois2023medrec}. These methods improve the accuracy and safety of recommendations by enhancing medication representations. However, the representation of diagnoses and procedures, along with their hierarchical structures, remains underexplored.
%

In medication recommendations, 
patient heterogeneity introduces a wide range of diagnoses and procedures, making it difficult for the model to effectively learn their feature representations from scratch.
For example, in the clinical dataset \text{MIMIC-III}~\cite{johnson2016mimic}, there are only 5,449 patients with at least two visits, yet exhibit 4,491 unique diagnoses and 1,412 unique procedures. 
%
On the other hand, the current models employ pooling operations (e.g., sum-pooling) to aggregate multiple diagnosis/procedure features as patient representation. Then, the model is optimized to match patients with medications, overlooking the structural properties (hierarchies) of diagnosis and procedure. We present the theoretical analysis in the following section. Due to the above issues, this paper leverages prior hierarchies to construct hierarchical representations rather than learning their complicated relationships from data. 

In clinical practice, it is a well-recognized principle that patients with similar diagnoses are often prescribed similar medications. Motivated by this, we highlight the importance of the hierarchical structures of diagnoses and procedures and exploit them to obtain more accurate representations. Our approach is grounded in the use of standardized medical codes, in which diagnoses and procedures exhibit a tree-like structure (hierarchical structure) from general to specific, corresponding to their conceptions. To be more specific, each diagnosis and procedure is tagged by a unique code, i.e., ICD coding\footnote{The International Classification of Diseases (ICD) is an international classification system used by healthcare providers to code and categorize diseases, conditions, etc.} in EHRs. 
However, current methods map diagnoses and procedures directly to one-hot encodings, which unfortunately disregard their hierarchical structure. 

To bridge the research gap, we propose the \underline{Hier}archical encoder named \textbf{HIER} for diagnosis and procedure representations guided by medical codes. We explicitly equip the model with hierarchical structures to enhance their representations and improve recommendation accuracy. 
Specifically, the proposed hierarchical encoder consists of two components: \textit{Relation Embedding} and \textit{Position Encoding}. The relation embedding aggregates their parental categories and is optimized with a self-supervised objective to capture neighbor hierarchical structure. The position encoding, on the other hand, is derived from segmented medical codes to introduce global hierarchical positions. Integrating these two components enables us to obtain hierarchical representations that are compatible with any medication recommendation model.
The main contributions of this paper are as follows:

\begin{itemize}
    \item We demonstrate, for the first time, that state-of-the-art medication recommendation models fail to adequately capture the structural relations of diagnoses and procedures. 
    \item We introduce a novel hierarchical encoder, guided by medical codes, to integrate hierarchical structure into the representation of medical terms (e.g., diagnoses and procedures). This hierarchical representation is universal and independent of the training process. 
    \item We conduct comprehensive experiments to validate the effectiveness of our method, demonstrating its significant and consistent performance improvements across four baselines and two real-world datasets.
\end{itemize}


\section{Related Works}
\subsection{Medication Recommendation}
The medication recommendation system provides the most suitable medications based on a patient's health conditions. Based on the input of single or multiple visits, medication recommendation can be categorized into instance-based~\cite{zhang2017leap,gong2021smr} and longitudinal methods~\cite{safedrug,tois2023medrec}.
Early studies are instance-based methods~\cite{zhang2017leap,wang2019order} and can only prescribe on single-visit patient data. LEAP~\cite{zhang2017leap} formulates the medication recommendation as a multi-instance, multi-label learning problem, mapping a set of instances, i.e., disease conditions to the associated medication sets. CompNet~\cite{wang2019order} formulates the task as an order-free Markov decision process, employing deep Q-Learning to capture correlations between medicines.
 
Recently, the longitudinal methods have attracted growing research interest~\cite{shang2019gamenet,tois2023medrec,yang2023molerec}. Longitudinal-based methods make use of the temporal dependencies among multiple visits in patients’ medical history. RETAIN~\cite{choi2016retain} is a two-level attention-based model that detects influential past visits and significant clinical variables within those visits. GAMENet~\cite{shang2019gamenet} proposes a graph-augmented memory module to store historical drug information as references for further prediction. SafeDrug~\cite{safedrug} extracts and encodes drug molecule structure to correlate with its functionalities and improve medication recommendation safety. MICRON~\cite{ijcai2021MICRON} adopt a residual-based recurrent network and takes the changes in patient health records as input to update a hidden medication vector. MoleRec~\cite{yang2023molerec} leverage a molecular substructure-aware attentive method to model substructures’ interactions and relevancy to patient’s health condition.
However, these methods do not target the representation of medical diagnoses and procedures, leaving much room for improvement.
 
\subsection{The Representation of Medical Codes}
ICD coding is a medical glossary established by the World Health Organization that provides specific coding rules and descriptions for diagnoses and procedures. 
%
Recently, several works have exploited the hierarchy of ICD codes for health applications, mainly focusing on automated ICD coding~\cite{wang2022hienet,kaur2023ai,teng2024few} and disease prediction\cite{choi2017gram,placido2023deep}.
The hierarchical information of ICD codes reduces the semantic searching space of predictions. 
GRAM~\cite{choi2017gram} employs a weighted average of the ancestors of a medical concept for diagnosis prediction but requires an expanded embedding matrix, which places a significant burden on training.
\cite{cikm2019ehr} leverages graph convolutional network (GCN) to capture the hierarchical relationships among medical codes and their semantics. They employ a label-dependent attention mechanism, allowing the model to learn distinct clinical document representations for predicting diagnosis or procedure codes.
HieNet~\cite{wang2022hienet} proposes a bidirectional hierarchy passage encoder to capture the codes’ hierarchical representations and a progressive predicting method for automated ICD coding. 

Notably, these end-to-end methods cannot be applied to medication recommendation systems. 
To bridge this gap, we aim to devise a simple yet effective hierarchical encoder to integrate their hierarchy into the models.


\section{Preliminary}
\subsection{Electrical Health Records (EHR)}
EHR data is a collection of patients' medical histories in the format of medical codes (e.g., diagnoses, procedures, and medications). Formally, the EHR for patient $i$ can be represented as a sequence of visits: 
$ \mathbf{X}_i = [\mathbf{x}_i^{(1)}, \mathbf{x}_i^{(2)}, \ldots, \mathbf{x}_i^{(T_i)}]$ 
where $T_i$ is the total number of visits for patient $i$.
To minimize confusion, we will omit the patient index $i$ as long as it is clear. Each visit of patient can be represented by $\mathbf{x}^{(t)}=[\mathbf{d}^{(t)},\mathbf{p}^{(t)},\mathbf{m}^{(t)}]$ containing the corresponding 
diagnosis codes $\mathbf{d}^{(t)} = \{d_1, d_2, \ldots\} \in \mathbb{R}^\mathcal{|D|}$, 
procedure codes $\mathbf{p}^{(t)} = \{p_1, p_2, \ldots\} \in\mathbb{R}^\mathcal{|P|}$ and 
medication codes $\mathbf{m}^{(t)} = \{m_1, m_2, \ldots\} \in\mathbb{R}^\mathcal{|M|}$, where $\mathcal{D},\mathcal{P},\mathcal{M}$ are all possible code sets, and $|\cdot|$ denotes the cardinality. 

\subsection{Patient Representation}
In medication recommendation, a patient is typically represented based on their diagnoses and procedures  \cite{shang2019gamenet,safedrug,tois2023medrec}. 
Current methods transform each diagnosis/procedure into a unique one-hot encoding, i.e., ID, subsequently projecting it into embeddings using two learnable embedding tables $\mathbf{E}_d\in\mathbb{R}^{|\mathcal{D}|\times dim}$ and $\mathbf{E}_p\in\mathbb{R}^{|\mathcal{P}|\times dim}$ for diagnoses and procedures, respectively. The embedding dimension is denoted by $dim$. This process is formalized as follows:

\begin{eqnarray}
    \mathbf{d}^{(t)}_e = \mathbf{E}_d \cdot \mathbf{d}^{(t)} ,\quad 
    \mathbf{p}^{(t)}_e = \mathbf{E}_p \cdot \mathbf{p}^{(t)}
\end{eqnarray}
where $\mathbf{d}^{(t)}$ and $\mathbf{p}^{(t)}$ are the diagnosis and procedure sets (multi-hot vectors) for the visit $t$ of a patient; $\mathbf{d}_e^{(t)}$ and $\mathbf{p}_e^{(t)}$ are the corresponding embedding sets with variable-size.

Generally, a patient has multiple diagnoses and multiple procedures in a visit. 
Therefore, their embeddings are aggregated using a pooling operator $\phi(\cdot)$, such as sum-pooling and average-pooling. The role of pooling is to merge a collection of diagnoses/procedure features into a single summary feature, while ensuring permutation invariance to input features. It can be formalized as follows: 
\begin{eqnarray}
    \mathbf{e}^{(t)}_d = \phi(\mathbf{d}^{(t)}_e) ,\quad  
    \mathbf{e}^{(t)}_p = \phi(\mathbf{p}^{(t)}_e)
\end{eqnarray}
where $\mathbf{e}^{(t)}_d$, $\mathbf{e}^{(t)}_p \in \mathbb{R}^{dim}$  are the aggregated diagnosis and procedure embeddings. They maintain a constant dimensionality across visit $t$.

For representing longitudinal patient data, two distinct Recurrent Neural Networks (RNNs) ~\cite{elman1990rnn} are employed to separately model the dynamic histories of patient diagnoses and procedures in different visits.
\begin{equation}
    \begin{aligned}
    \mathbf{h}^{(t)}_d = \text{RNN}_d(\mathbf{e}^{(t)}_d,\mathbf{h}^{(t-1)}_d)  
    ,\quad
    \mathbf{h}^{(t)}_p = \text{RNN}_p(\mathbf{e}^{(t)}_p,\mathbf{h}^{(t-1)}_p) 
    \end{aligned}
\end{equation}
where $\mathbf{h}^{(t)}_d$, $\mathbf{h}^{(t)}_p \in \mathbb{R}^{dim}$ are the hidden states of RNNs. The hidden states of $\mathbf{h}^{(0)}_d$ and $\mathbf{h}^{(t)}_p$ are initialize from zero.

Finally, we concatenate the two hidden states to generate the comprehensive patient representation $\mathbf{e}_{pt}^{(t)}$.
\begin{equation}
    \mathbf{e}_{pt}^{(t)} = Concate(\mathbf{h}^{(t)}_d, \mathbf{h}^{(t)}_p)
\end{equation}
As shown in Fig. \ref{fig:model}(a), the patient representation is then processed through various model backbones to predict medication combinations (in this paper, we refer to the remaining modules of the model as the backbone). We do not show RNN in Fig. \ref{fig:model}, because it is unnecessary for the instance-based medication recommendation methods. 

\subsection{International Classification of Diseases (ICD)}
%
ICD provides a comprehensive list of codes for diseases, conditions, injuries, surgeries, and other medical procedures. We take extensively used ICD-9-CM
\footnote{https://www.cms.gov/medicare/coding-billing/icd-10-codes/icd-9-cm-diagnosis-procedure-codes-abbreviated-and-full-code-titles} 
as an example, which is an adaptation of ICD-9 specifically for use in the United States. Other versions of ICD codes follow a similar hierarchical structure. In ICD-9-CM, the three-digit codes are the heading for a category that can be further subdivided into more specific codes using fourth and/or fifth digits. We illustrate the detailed code structure for ``Diabetes mellitus'' as follows:


\begin{equation}
    \underbrace{\boxed{2} \boxed{5} \boxed{0}}_{\text{Category}}. \underbrace{\boxed{0} \boxed{1}}_{\text{Etiology}}
\end{equation}

\begin{itemize}
    \item \textbf{250} \textit{Diabetes mellitus}
    \begin{itemize}
        \item \textbf{250.0} 
        \textit{Diabetes mellitus without mention of complication}
        \begin{itemize}
            \item \textbf{250.01} \textit{Diabetes mellitus without mention of complication, type I [juvenile type], not stated as uncontrolled}
        \end{itemize}
    \end{itemize}
\end{itemize}
where {\boxed{2}\boxed{5}\boxed{0}} categorizes ``Diabetes mellitus''. The fourth and fifth digits of \boxed{0}\boxed{1} provide additional code notes for the etiology (manifestation) specific to the particular diabetic manifestation. Additionally, different code ranges are meaningful. For instance, \text{249-259} and \text{240-279} stand for ``Diseases of other endocrine glands'' and ``Endocrine, nutritional and metabolic diseases, and immunity disorders''.


\section{Proposed Method}
\begin{figure}[htbp]
    \centering
    \includegraphics[width=1.0\textwidth]{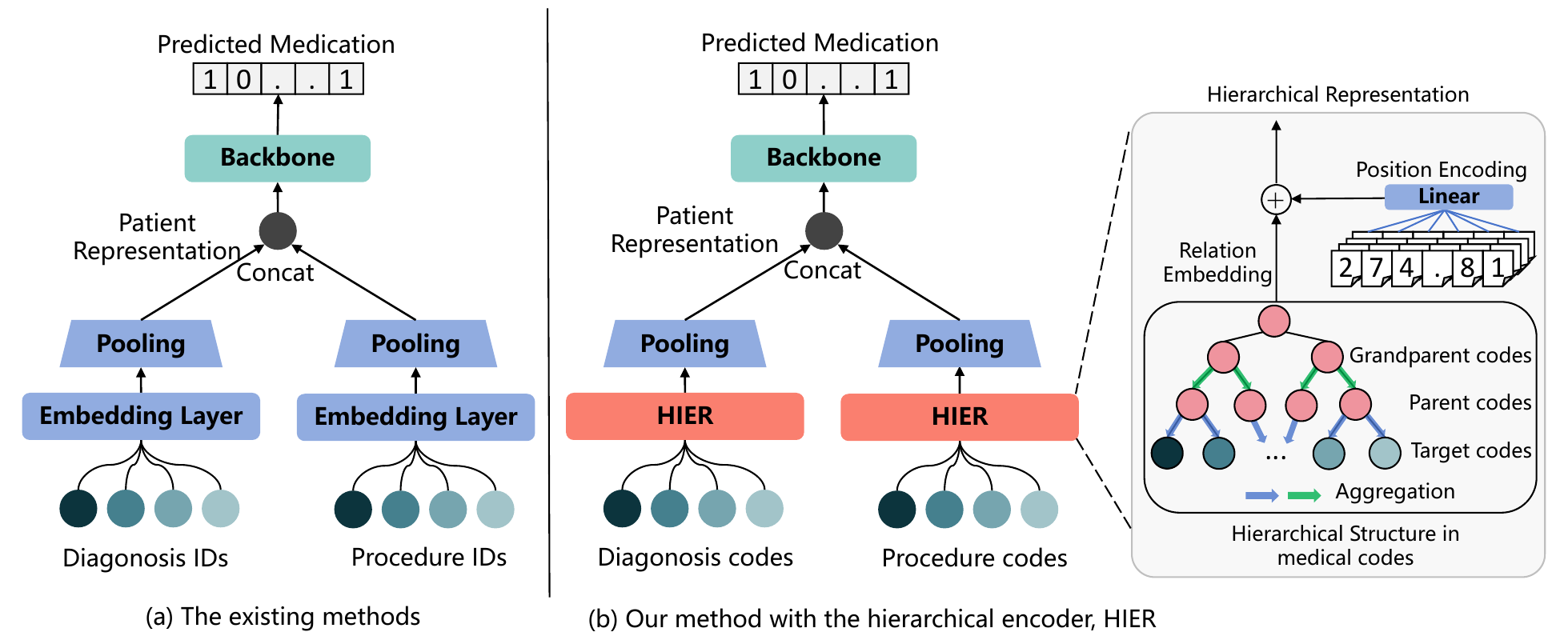}
    \caption{An illustration of medication recommendation models, where (a) is the paradigm of existing methods; (b) is our method with the hierarchical encoder.}
    \label{fig:model}
\end{figure}

In this section, we introduce a simple yet effective \underline{hier}archical encoder, \textbf{HIER}, to incorporate the hierarchical structures based on medical codes and construct the hierarchical representation. As illustrated in Fig. \ref{fig:model}(b), in terms of design, our method can seamlessly integrate into the existing methods by replacing the traditional embedding layer. By doing so, we can effectively enhance feature representations, improving recommendation accuracy.

\vspace{-10pt}
\subsection{Hierarchical Encoder}
This paper targets the medical code of diagnoses and procedures, but our method can also be used in other tree-structured terms. Without loss of generality, we refer to them as medical codes. As illustrated in Fig. \ref{fig:model}(b), the proposed hierarchical encoder first aggregates the target medical code with higher-level codes (e.g., parent codes) to generate relation embedding. It can establish neighbor hierarchical relationships in the embedding space. Afterward, the position encoding introduces the global hierarchical position from the separated segments of codes. Finally, we obtain the hierarchical representation by merging the relation embedding with the position encoding.

\subsubsection{Relation Embedding}
The relation embedding establishes relative hierarchical relationships in diagnosis and procedure embeddings.
Firstly, we transform the medical coding system, such as ICD-9-CM, into two undirected tree-structured graphs for diagnoses and procedures, respectively. This structure describes the hierarchy (topology) among the medical codes, uniformly denoted by $T=(\mathcal{V}, \mathcal{E})$.

Then, we encode the target medical code $v$ by aggregating its embedding with its parent codes denoted by $p(v)$ within the tree. For instance, the diagnosis code \text{250.01} can be derived through the sequence of higher-level codes (grandparent and parent codes) \text{250} $\to$ \text{250.0} $\to$ \text{250.01}. This aggregation from higher-level codes provides a relative hierarchical connection between the target code and its ancestors. 
We formalize the aggregation process as follows:
\begin{align}
    \mathbf{e}_v^{k+1} 
    &= Aggregator\left( {\mathbf{e}_v^k , \mathbf{e}_{p(v)}^k} \right) \\
    &= \sigma \left( \mathbf{W}^{k+1} \cdot Concat\left( {\mathbf{e}_v^k, \mathbf{e}_{p(v)}^k} \right) \right)
\end{align}
where $\mathbf{e}_v^{k+1}\in \mathbb{R}^{dim}$ is the aggregated embedding of ${v}$ at layer $k+1$; $\mathbf{e}_v^k$, $\mathbf{e}_{p(v)}^{k} \in \mathbb{R}^{dim}$ are the embeddings of $v$ and its parent $p(v)$ at layer $k$, respectively. $\sigma$ is the rectified linear unit (ReLU) activation function. $\mathbf{W}^{k+1}\in \mathbb{R}^{2dim\times dim}$ is the weight matrix.

By aggregating $K$ layers, we obtain the relation embedding for the medical code. We randomly initialize each node embedding at $k=0$ and optimize the weight matrix with a self-supervised objective (which will be introduced later) to preserve the hierarchical structure in the embedding space. Finally, the derived relation embedding can be used in various downstream tasks, such as medication recommendation. It can be formulated as follows:
\begin{equation}
    \mathbf{r}_v \gets  \mathbf{e}_v^{K},\forall v \in \mathcal{V}
    \label{eq:final_hier_emb}
\end{equation}
where $\mathbf{r}_v \in \mathbb{R}^{dim}$ is the relation embedding for the medical code $v$, after $K$ layers of aggregation .


\subsubsection{Position Encoding}
The above relation embedding provides hierarchical information on the neighbor structure. To incorporate the global hierarchical feature, we devise a hierarchical position encoding for medical codes. 
Specifically, we first separate the medical code into a sequence of digits. Let $\{v_1, \dots, v_L\}$ represent the sequence of digits of the medical code $v$, where $L$ is the number of digits. Then, each digit $v_l$ is transformed to one-hot encoding.
Finally, we concatenate all one-hot encodings and obtain the multi-hot encoding.
\begin{eqnarray}
\text{PE}_v = \left[ \text{PE}_{(v_1)}, \text{PE}_{(v_2)}, \dots, \text{PE}_{(v_L)} \right]
\end{eqnarray}
where $\text{PE}_{(v_l)}$ is the one-hot encoding of the $l$-th digit in the medical code $v$; $\text{PE}_v$ is multi-hot position encoding. For the short code, zero padding is applied to the missing digits to maintain dimensional consistency.
To adjust the position encoding to a specific dimension, we adopt the linear layer followed by a ReLU function $\sigma$. The ReLU activation function ensures that the projected position encoding remains positive, preventing symbol conflicts when facing the pooling operator. We denote the final position encoding as $\mathbf{s}_v$. 
\begin{equation}
    \mathbf{s}_v = \sigma(Linear(\text{PE}_v))
    \label{eq:hier_pos}
\end{equation}
By adding the relation embedding $\mathbf{r}_v$ with the position encoding $\mathbf{s}_v$, we obtain the hierarchical representation of medical codes $\mathbf{z}_v \in \mathbb{R}^{dim}$.
\begin{equation}
    \mathbf{z}_v \gets \mathbf{r}_v + \mathbf{s}_v
\end{equation}

\subsection{Implementation}
In order to seamlessly integrate our method into various SOTA medication recommendation backbones, we propose a simple and efficient hierarchical encoder plug-in to replace the traditional embedding layer, as shown in Fig. \ref{fig:model}. The relation embedding is universal and generated independently from the model training, ensuring minimal additional computational and parameter overhead.

%
Specifically, we optimize the relation embedding in a self-supervised manner to preserve the hierarchical structure. We encourage the relation embedding $\mathbf{r}_v$ of medical code $v$ to be closer to the embedding of its parent code $\mathbf{r}_{p(v)}$ and farther away from the randomly selected one $\mathbf{r}_{rand}$. The self-supervision optimization objective is formulated as follows:
\begin{equation}
    \mathcal{L} = -\frac{1}{N} \sum_{n=1}^{N} \left[ \log(\mathbf{r}_v \cdot \mathbf{r}_{p(v)}) + \log(\mathbf{r}_v \cdot \mathbf{r}_{rand}) \right]
\end{equation}
where $N$ is the number of samples in the batch; $\mathbf{r}_{p(v)}$ and $\mathbf{r}_{rand}$ are the relation embedding of its parent and randomly selected code. 

Next, we generate the position encoding $\mathbf{s}_v$ as previously described, and add the relation embedding $\mathbf{r}_v$ to obtain the hierarchical representation $\mathbf{z}_v \gets \mathbf{r}_v + \mathbf{s}_v$. During training for specific tasks, we only adjust the parameters of derived relation embedding and the weight matrix of position encoding.


\subsection{Theoretical Insights}
In this subsection, we present some theoretical insights to show the limitations of existing methods. We analyze from two perspectives: \textit{Alignment and Uniformity}, as well as \textit{Feature Aggregation}. They can address two crucial questions:

\subsubsection{What knowledge do existing models derive from the data?} 

Inspired by \cite{wang2020understanding}, we analyze the Binary Cross-Entropy (BCE) loss, a widely used loss function in medication recommendation. We derive that the BCE loss favors the two properties of alignment and uniformity from Eq.(\ref{eq:bce}) to Eq.(\ref{eq:ali&uni}). For the proof, please see the Appendix.

\begin{align}
\mathcal{L}_{\text{BCE}} 
&= -\frac{1}{N}\sum_{i=1}^{N} \left[ y_i \log(\sigma(\hat{y}_i)) + (1-y_i) \log(1-\sigma(\hat{y}_i)) \right] \label{eq:bce}\\
&= - \frac{1}{N}\sum_{i=1}^{N}[ \underbrace{y_i f(u_i)^\top f(m_i)}_{\text{Alignment}} - \underbrace{\log(1+e^{f(u_i)^\top f(m_i)})}_{\text{Uniformity}}]  \label{eq:ali&uni}
\end{align}
where $y_i$ and $\hat{y}_i$ represent the label and prediction for sample $i$, respectively; $\sigma$ is the activation function and $N$ denotes the number of batch samples. 

Assuming $u$ is the patient and $m$ is the medication, the neural network $f(\cdot)$ maps them into a low-dimensional embedding $f(u)$ and $f(m)$, respectively. If the matching score is determined through the dot product $f(u)^\top f(m)$, then we can have Eq. (\ref{eq:ali&uni}), where the optimization objective includes both Alignment and Uniformity terms. 

The first term encourages the alignment of paired patient $u_i$ and medication $m_i$. If $y_i=1$, it maximizes the inner product of $f(u_i)$ and $f(m_i)$ . Consequently, the matched patient and medication features align closely. The second term preserves maximal information by the uniform distribution of patient and medication features on the unit hypersphere~\cite{wang2020understanding,wang2022towards}.
In summary, the existing models essentially optimize the objectives in terms of Alignment and Uniformity. That is, it matches the patient with possible medications while uniformly distributing the features of patients and medications. 

\subsubsection{Why do existing models fail to capture hierarchical relationships?}
In medication recommendation, the model aggregates multiple diagnosis/procedure features with the pooling operator $\phi$ (e.g., sum-pooling), which also makes neural networks permutation-invariant for the features.  \cite{zaheer2017deepset} shows that the pooling operator encodes the entire set (i.e., multiple features), and then the model backbone, in principle, recovers all the input information and performs relational reasoning from there. However, any such relational reasoning is not built into the model, and must be entirely learned, leading to an additional training burden~\cite{murphy2018janossy,Zhang2019FSPool,wagstaff2022universal}. Hence, the model focuses more on individual attributes of the inputs and less on encoding information about their interactions and relationships.

In our case, the model struggles to adequately learn the relations of diagnoses and procedures, particularly their hierarchical relations. Therefore, in this paper, we explicitly incorporate the hierarchical structure of the medical codes rather than relying solely on learning from data. 




\section {Experiment}
To verify the superiority and effectiveness of the proposed HIER method, we perform extensive
experiments to answer the following research questions:
\begin{itemize}
    \item \textbf{RQ1}: How does our method perform compared to various SOTA baselines on different datasets?
    \item \textbf{RQ2}: How do individual components affect the model performance?
    \item \textbf{RQ3}: How does our method improve convergence efficiency during training?
    \item \textbf{RQ4}: Do the traditional embedding method and our approach capture hierarchical relationships?

\end{itemize}

\subsection{Experimental Setup}
The models are trained on Ubuntu 22.04 with NVIDIA RTX 3090. The implementation is based on the Pyhealth framework~\cite{pyhealth2023yang}. We follow most of the hyper-parameter configures of the original baselines. We set the learning rate to $\{10^{-3}, 10^{-4}, 10^{-5}\}$, and set the epochs to 25 (sufficient to achieve optimal performance).
Following the recent studies~\cite{safedrug,yang2023molerec,tois2023medrec}, we evaluate the model using Jaccard Similarity Score (Jaccard), F1 score, Precision-Recall AUC (PRAUC), and Drug-Drug Interaction rate (DDI). A lower DDI indicates better performance, while higher values are better for other metrics.

\subsubsection{Datasets}
We perform the experiments on two extensively used datasets {MIMIC-III}~\cite{johnson2016mimic} and {MIMIC-IV}~\cite{johnson2023mimic}. The statistics of the pre-processed data are reported in Tab. \ref{tab:statistics}. Following the pre-processing of \cite{shang2019gamenet,safedrug}, we filter out the patients with only one visit, as they do not reflect long-term health status and medication response. 
It is worth noting that we retain tail diagnoses and procedures for a more realistic clinical setting. 
We split the patients into training, validation, and test sets with a ratio of 8:1:1.



\subsubsection{Baselines}
We consider the following popular and advanced baseline methods for medication recommendation. For a fair comparison, our method only modifies the baseline model's embedding layers with HIER. 
Notably, there is no directly available hierarchical representation method for medication recommendation, as related works are end-to-end models for other medical applications.

\begin{itemize}
    \item RETAIN~\cite{choi2016retain} uses a two-level attention mechanism to detect influential past visits and significant clinical variables within those visits, e.g., key diagnoses. 
    \item GAMENet~\cite{shang2019gamenet} propose graph augmented memory networks to encode longitudinal patient data and drug-drug interaction graphs.
    \item SafeDrug~\cite{safedrug} leverages drug molecule structures and DDI graph to recommend effective and safe medication combinations.
    \item MoleRec~\cite{yang2023molerec} propose a substructure-aware attentive method that models substructures’ interactions and relevancy to a patient’s health condition.
\end{itemize}

\begin{table}[!h]
    \centering
    \caption{The statistics of \text{MIMIC-III} and \text{MIMIC-IV} datasets.}
    \scalebox{1}{
    \begin{tabular}{l|rr}
    \toprule
        \textbf{Item} &  \textbf{MIMIC-III}  & \textbf{MIMIC-IV}  \\
        \midrule
        \# of visits / \# of patients 
        & 14,141 / 5,449        & 147,406 / 43,939 \\
        diag. / proc. / med. space size 
        & 4,491 / 1,412 / 193   & 19,184 / 10,603 / 200 \\
        avg. / max \# of visits 
        & 2.59 / 29             & 3.35 / 70  \\
        avg. / max \# of diag. per visit 
        & 31.65 / 459           & 48.98 / 1,318\\
        avg. / max \# of proc. per visit 
        & 9.46 / 85             & 8.763 / 164\\
        avg. / max \# of med. per visit 
        & 29.74 / 105           & 23.83 / 111\\
        \bottomrule
    \end{tabular}
    }
    \label{tab:statistics}
\end{table}

\subsection{Performance Comparison (RQ1)}


\begin{table}[!ht]
\centering
\caption{Performance comparison with four baseline methods across two datasets.}
\scalebox{0.76}
{
\begin{tabular}{cccccccccc}
\toprule
    \multicolumn{2}{c}{\multirow{2}{*}{Methods}} & \multicolumn{4}{c}{MIMIC-III} & \multicolumn{4}{c}{MIMIC-IV} \\
    \cmidrule{3-10}  
    \multicolumn{2}{c}{}      & Jaccard$(\uparrow)$  & F1-score$(\uparrow)$  & PRAUC$(\uparrow)$ & DDI$(\downarrow)$ & Jaccard$(\uparrow)$  & F1-score$(\uparrow)$ & PRAUC$(\uparrow)$ & DDI$(\downarrow)$ \\
    \midrule

    \multirow{2}{*}{RETAIN~\cite{choi2016retain}} 
    & & 0.4645 & 0.6224  & 0.7376  & 0.0680 
    & 0.4673 & 0.6197 & 0.7390 & \textbf{0.0629}\\ 
    & +Ours \cellcolor{mygray} 
    & \textbf{0.4837} & \textbf{0.6405} & \textbf{0.7540} & \textbf{0.0660}   
    & \textbf{0.4710} & \textbf{0.6236} & \textbf{0.7438} & \text{0.0632} \\ 

    \multirow{2}{*}{GAMENet~\cite{shang2019gamenet}} 
    & & 0.4485 & 0.6075  &  0.7136  &  0.0679
    & 0.4544 & 0.6073 & 0.7173 & 0.0652 \\ 
    & +Ours  \cellcolor{mygray} 
    & \textbf{0.4739} & \textbf{0.6307} & \textbf{0.7402} & \textbf{0.0663}  
    & \textbf{0.4777} & \textbf{0.6305} & \textbf{0.7506} & \textbf{0.0639} \\ 

    \multirow{2}{*}{SafeDrug~\cite{safedrug}}  
    & & 0.4541 &  0.6132  & 0.7174  & 0.0710 
    & 0.4519 & 0.6038 & 0.7121 & 0.0638 \\
    & +Ours \cellcolor{mygray} 
    & \textbf{0.4745} & \textbf{0.6314} & \textbf{0.7405} & \textbf{0.0692} 
    & \textbf{0.4704} & \textbf{0.6219} & \textbf{0.7420} & \textbf{0.0614}\\


    \multirow{2}{*}{MoleRec~\cite{yang2023molerec}}  
    &
    & 0.4572 & 0.6159 & 0.7157 & 0.0684
    & 0.4302 & 0.5865 & 0.6962 & 0.0642
    \\
    &+Ours \cellcolor{mygray} 
    & \textbf{0.4627} & \textbf{0.6213} & \textbf{0.7252} & \textbf{0.0652}
    & \textbf{0.4324} & \textbf{0.5904} & \textbf{0.7000} & \textbf{0.0641}\\
    \bottomrule 
\end{tabular}
}
\label{tab:perf}
\end{table}

As shown in Tab. \ref{tab:perf}, we evaluate the performance of our hierarchical encoder HIER with four baseline methods (RETAIN, GAMENet, SafeDrug and MoleRec).
We observe that our method consistently improves performance across all baselines in terms of Jaccard, F1 and PRAUC in both datasets. For example, our method improves GAMENet by 5.66\%, 3.82\% and 3.73\% in Jaccard, F1, and PRAUC, respectively, on the MIMIC-III dataset. This is attributed to the fact that our method effectively improves the representations of diagnoses and procedures by incorporating their hierarchies.

We also find that HIER reduces DDIs in certain cases, except for similar results compared to RETAIN and MoleRec on MIMIC-IV. While our method does not specifically target drug-drug interactions, the results suggest that improving the representations of diagnoses and procedures can help reduce DDIs.

It is noteworthy that the classic model, RETAIN, achieves the best performance except for DDIs than other baselines on both datasets. For example, in the MIMIC-III dataset, RETAIN achieves the best result of PRAUC 0.7376, while GAMENet, SafeDrug, and MoleRec achieve only 0.7136, 0.7174, and 0.7157, respectively. This is due to the realistic and challenging experimental setup that does not filter out tail diagnoses and procedures. RETAIN effectively predicts medications by identifying the importance of visits and the conditions of patients. The simple architecture of RETAIN is more effective than others. In addition, MoleRec's performance collapses in MIMIC-IV. This may be because the model fails to correlate the fine-grained substructures of drug molecules with spare and diverse diagnoses and procedures in this dataset. This phenomenon inspires us to explore more effective model architectures for future work.


\subsection{Ablation Study (RQ2)}
We conduct an ablation study for the relation embedding and the position encoding to evaluate their effectiveness. The experiments are conducted on the MIMIC-III dataset with the baseline of SafeDrug. We have two variants when only using relation embedding or position encoding denoted by `\emph{w/} E' and `\emph{w/} P', respectively. As shown in Table. \ref{tab:ablation}, we observe that significant performance gains can still be achieved by using either relation embedding or position encoding. The position encoding can perform better than the relation embedding, thanks to its global hierarchical position. These results demonstrate the effectiveness of hierarchical information. Finally, our method obtained the best results on all metrics with the combination of the two components.

\begin{figure}[!tbp]
    \centering
    \scalebox{0.3}{
    \includegraphics{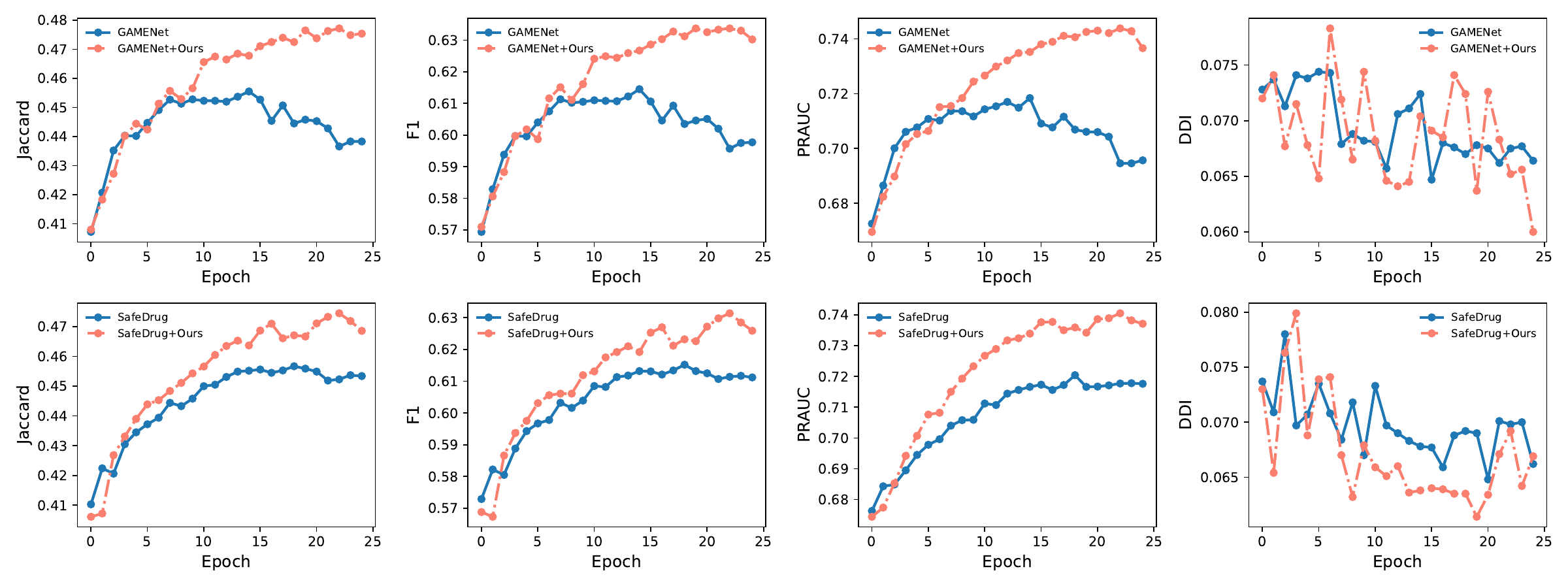}
    }
    \caption{Convergence efficiency curve in the test set during training stage: the first row for GAMENet and the second row for SafeDrug. Our method significantly improves the training convergence efficiency in both baselines. 
    }
    \label{fig:convergence}
\end{figure}

\begin{table}[h]
    \centering
    \caption{Ablation study on MIMIC-III dataset. The models with relation embedding or position encoding are denoted by `\emph{w/} E' and `\emph{w/} P', respectively.}
    \begin{tabular}{c|cccc}
    \toprule
        Model &  Jaccard$(\uparrow)$  & F1-score$(\uparrow)$  & PRAUC$(\uparrow)$ & DDI$(\downarrow)$ \\
        \midrule
        \emph{w/o} E\&P & 0.4541 & 0.6132 & 0.7174 & 0.0710\\
        \emph{w/} E & 0.4668  & 0.6245 & 0.7346 & 0.0698\\
        \emph{w/} P & 0.4717  & 0.6292 & 0.7368 & 0.0727 \\ 
        \midrule
        \textbf{Ours}  & \textbf{0.4745} & \textbf{0.6314} & \textbf{0.7405} & \textbf{0.0692} \\
        \bottomrule
    \end{tabular}
    \label{tab:ablation}
\end{table}

\subsection{Convergence Efficiency (RQ3)}
We also analyze the training convergence efficiency in comparison with two baselines, GAMENet and SafeDrug, as shown in Fig.\ref{fig:convergence}. We evaluate the performance on the test set during the training stage every epoch (maximum of 25). 
As training progresses, both GAMENet+Ours and SafeDrug+Ours can surpass the baseline models in terms of Jaccard, F1, and PRAUC. The baseline models (GAMENet and SafeDrug) overfit or stabilize around epoch 15. In contrast, the performance of our method continues to improve and has not yet reached its optimum, suggesting an enhanced generalization capability.

The results illustrate that our method effectively accelerates training progress and improves the model's generalization ability. We also find that our method slightly reduces the DDI, even though we do not control it.

\begin{figure}[t!]
    \centering
    \scalebox{0.45}{
    \includegraphics{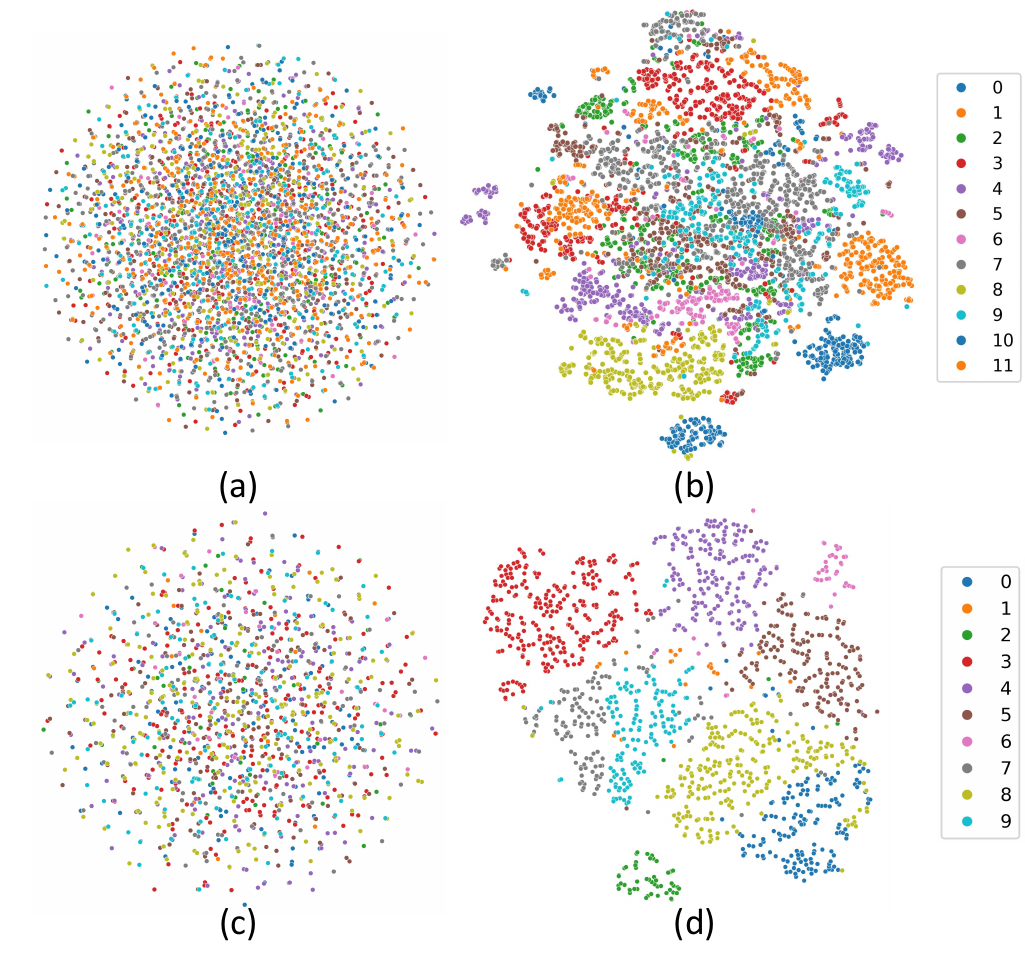}
    }
    \caption{The visualization of diagnoses and procedures in the 2-d space using t-SNE, where (a) and (c) for embeddings of diagnoses and procedures; (b) and (d) for our relation embeddings of diagnoses and procedures.}
    \label{fig:tsne}
\end{figure}

\subsection{Visualization and Analysis (RQ4)}
The goal of our method is to incorporate the hierarchical structure into the representation of diagnoses and procedures via medical codes. To validate it, we employ the t-SNE~\cite{t-SNE} tool to visualize the traditional embeddings and relation embeddings of diagnoses and procedures on a 2-dimensional space. As illustrated in Fig.~\ref{fig:tsne}, we visualize the diagnosis and procedure embeddings on the trained models SafeDrug and SafeDrug+Ours. The different colors of points represent the main medical category, corresponding to the first digit of medical codes (including supplementary classification "E000-E999" and "V01-V91"). 

As shown in Fig. \ref{fig:tsne}(a)(c), the traditional embedding method exhibits a uniform distribution for both diagnoses and procedures. Their underlying relationships have not been captured. This is because the traditional embedding method fails to learn the hierarchical structure and treats diagnoses/procedures as separate and unrelated. In contrast, the results of our method in Fig. \ref{fig:tsne}(b)(d) show distinct clusters of relation embeddings corresponding to their hierarchical structure of medical codes. 
The visualization results demonstrate the effectiveness and superiority of our method through the hierarchical structure of medical codes.

\section{Conclusion and Future Work}
In this paper, we proposed a simple yet efficient hierarchical encoder plug-in for medical codes in medication recommendation. It can effectively incorporate the hierarchical structure of diagnoses and procedures into their representation and improve recommendation accuracy.
In future work, we intend to investigate the applicability of our hierarchical encoder across different healthcare applications. In addition, we hope to evaluate the model in diverse clinical settings to understand its generalizability and robustness.

\newpage


%
%



%
%
%
\bibliographystyle{splncs04}
\bibliography{ref}
\newpage
\appendix
\section{Appendix}
\vspace{-5pt}
\textbf{Proof of Equation 14}:

We present proof of the loss of Binary Cross Entropy (BCE) in Alignment and Uniformity.
Assuming $u$ is the patient and $m$ is the medication, the neural network $f(\cdot)$ maps them into a low-dimensional representation $f(u)$ and $f(m)$, respectively. If the matching score is determined through the dot product $f(u)^\top f(m)$, then the loss can be described as:
\begin{align}
\mathcal{L}_{BCE} &= -\frac{1}{N} \sum_{i=1}^{N} \left[ y_i log(\sigma(\hat{y}_i)) + (1-y_i)log(1-\sigma(\hat{y}_i)) \right] \notag\\
&= -\frac{1}{N} \sum_{i=1}^{N} \left[ y_i log(\sigma(f(u_i)^\top f(m_i))) + (1-y_i) log(1-\sigma(f(u_i)^\top f(m_i))) \right] \notag\\
&=-\frac{1}{N}\sum_{i=1}^{N}\left [ y_i log(\frac{1}{1+e^{-f(u_i)^\top f(m_i)}}) + (1-y_i) log(1-\frac{1}{1+e^{-f(u_i)^\top f(m_i)}}) \right]  \notag
\end{align}
where $y_i$ and $\hat{y}_i$ represent the label and prediction for sample $i$, respectively; $\sigma$ is the Sigmoid function and $N$ denotes the number of samples. 

According to the law of logarithm, we have:
\begin{align*}
    log(\frac{1}{1+e^{-f(u_i)^\top f(m_i)}}) &= f(u_i)^\top f(m_i)-log(1+e^{f(u_i)^\top f(v_i)}) \\
    log(1-\frac{1}{1+e^{-f(u_i)^\top f(m_i)}}) &= -log(1+e^{f(u_i)^\top f(m_i)})
\end{align*}

Then, we substitute into the BCE loss to obtain: 
\begin{align}
\mathcal{L}_{BCE} = 
&= - \frac{1}{N}\sum_{i=1}^{N}\left [ y_i \left(f(u_i)^\top f(m_i) - log(1+e^{f(u_i)^\top f(m_i)})\right)  +  \left ( 1-y_i \right )\left ( -log(1+e^{f(u_i)^\top f(m_i)}) \right )    \right] \notag\\
&= - \frac{1}{N}\sum_{i=1}^{N}\left [ y_i f(u_i)^\top f(m_i) - y_i log(1+e^{f(u_i)^\top f(m_i)})  - \left ( 1-y_i \right )\left (log(1+e^{f(u_i)^\top f(m_i)}) \right ) \right] \notag\\
&= - \frac{1}{N}\sum_{i=1}^{N} \Big[ \underbrace{y_i f(u_i)^\top f(m_i)}_{\text{Alignment}} - \underbrace{\log(1+e^{f(u_i)^\top f(m_i)})}_{\text{Uniformity}}\Big] \notag
\end{align}

The optimization objective includes both alignment and uniformity. The first term encourages the alignment of paired patient$u_i$ and medication $m_i$. When $y_i=1$, it aims to make $f(u_i)$ and $f(m_i)$ similar by maximizing their inner product. Consequently, the matched patient and medication representation align closely.
The second term preserves maximal information by the uniform distribution of sample pairs on the unit hypersphere~\cite{wang2020understanding,wang2022towards}, which ensures the uniformity of the representations in the embedding space. It penalizes very high similarity scores indiscriminately, encouraging the model to spread out the embeddings more uniformly across the space. This term helps prevent the collapse of the embeddings into a trivial solution where all points are mapped to the same representation.


\end{document}